\documentclass[aps,pra,twocolumn,groupedaddress,showpacs]{revtex4}

\usepackage{graphicx}
\usepackage{dcolumn}
\usepackage{bm}
\usepackage{verbatim}
\usepackage{amsmath,amssymb}
\usepackage{relsize,exscale}
\usepackage{color}
\begin{document}

\title{ Replay to "Comment on 'Quantum phase for an arbitrary system with finite-dimensional Hilbert space'}

\author{Du\v san Arsenovi\'c $^1$, Nikola Buri\' c $^1${\thanks {buric@ipb.ac.rs}},\\
 Dragomir Davidovi\' c $^2$ and Slobodan Prvanovi\' c$^1$\\
$^1$  Institute of Physics, University of Belgrade, \\PO Box 68, 11000 Belgrade, Serbia.\\
$^2$ Vinca Institute of nuclear sciences, University of Belgrade, \\11001 Belgrade, P.O.Box
522, Serbia. }

\begin{abstract}
We point out the crucial difference between the relative and absolute phase observables treated in our contribution \cite{1} and in
 the Comment by Hall and Pegg \cite{HP} respectively. The main contribution of our work is to show that the quantum expectation of  the relative phase is  highly
    discontinuous function of the frequency and to point out interesting dependence of the phase on the number-theoretic nature of the frequencies.

\end{abstract}

\maketitle

The main difference between the object treated in our paper \cite{1} and  in it's   Comment by Hall and Pegg (HP)\cite{HP} is that we study
 what might be called relative phase (or relative age or angle) and HP analyze an absolute age.
 Consider a periodic system with the period $T$. Our relative phase measures the relative  part of the total period $T$
  undergone by the system during an interval $(0,t)$. In this way the relative ages of two systems with different periods can be
   meaningfully compared.
  It does not make sense to compare
  absolute ages of two systems with different periods (or life-spans), but it does make sense to compare their relative ages.
 There are processes performed by (or with) quantum systems (for example: STIRAP, quantum circuits etc...) such that the part of the process
  undergone by the system, as measured by the relative phase, and not the actual duration (as measured by the absolute phase), is the only
   relevant information. Such a process might take quite different time intervals to perform by different systems, but nevertheless one would consider two systems to be in the same phase of the process if the the relative phases are equal. Such comparison is the main motivation for the
    introduction of the relative phases.
   On the other hand, periodic systems with different frequencies can be used as clock's to measure the same absolute time.
   In this case
   the notion of an absolute phase, as a system's observable that measures the absolute time is appropriate. So, both relative and absolute phase can be defined and have meaningful interpretation.

  Our main motivation was to compare relative phases of different periodic systems, and to introduce a (relative) phase observable
   whose dependence on the frequencies can be meaningfully analyzed. In particular, we show that the quantum expectation of  such relative phase is  highly
    discontinuous function of the frequency and we point out interesting dependence on the number-theoretic nature of the frequencies.
    These properties of the relative phase as a function of the frequency are the main results of our short communication.

    Incommensurate eigenmodes imply in general quasi-periodic motion whose period can be formally considered as $T=\infty$. Such a system is
     equally old or young in any particular moment in time. In this sense our result for the relative phase (relative age, angle) of a quasi-periodic
      system is as it should be. We agree with the observation made by HP that it might be unnatural to rescale and represent a quasi-periodic motion  as dynamics on a
       unit circle. But this still does not mean that the relative age of a periodic and of a quasi-periodic system can not be meaningfully compared.
       Therefore, we compute explicitly the relative phase observable for periodic motion for all periods, which can and must be rescaled on the
        unit circle, and then consider the relative phase of the quasi-periodic motion as a limit of the corresponding  sequence
         of periodic relative phases.  No explicit representation of the quasi-periodic dynamics on a circle is involved in this process.

    The example provided by HP of a special initial condition leading to a periodic motion in a system with typically quasi-periodic orbits
    shows just the existence of such special initial conditions. Of course, they are not typical for the considered system with incommensurate
     eigenfrequencies  neither in the measure theoretic not in the topological sense.

     Discontinuous dependence of the relative phase on the frequency is the main topic of our communication. In particular we show
      how an $\epsilon$ perturbation of the frequency in general implies discontinuous changes of the relative phase. Our results, in fact show much more, i.e. we indicate special sequences of frequencies, namely those corresponding to the successive best rational approximants
      $p_n/q_n$ of an irrational $\nu$, over which the relative phase varies continuously and approaches the uniform distribution as $p_n/q_n\rightarrow \nu$.

       Finally, HP remark about lack of details in the presentation, in this  short communication, of our construction for the relative phase for countable and/or possibly degenerate spectrum. To this we can only answer that we are about to complete a longer paper with full theory and all such details.


  {\bf Acknowledgments} This work was supported in part by the Ministry
  of  Education and Science of the Republic of Serbia, under project No.
  171017, 171028 and 171006.


\begin{thebibliography}{99}
\bibitem{1}{D. Arsenovic, N. Buric, D. Davidovic and S. Prvanovic, Phys.Rev.A, {\bf 85}  044103 (2012).}

\bibitem{HP}{M. J. W. Hall, D. T. Pegg, Comment on "Quantum phase for an arbitrary system with finite-dimensional Hilbert space",   arXiv:1206.4385. }

\end{thebibliography}
\end{document}